\catcode`\@=11
\newif\if@fewtab\@fewtabtrue
{\count255=\time\divide\count255 by 60
\xdef\hourmin{\number\count255} 
\multiply\count255 by-60\advance\count255 by\time
\xdef\hourmin{\hourmin:\ifnum\count255<10 0\fi\the\count255}}
\def\ps@draft{\let\@mkboth\@gobbletwo
    \def\@oddfoot{\hbox to 7 cm{\tiny \versionno
       \hfil}\hskip -7cm\hfil\rm\thepage \hfil {\tiny\draftdate}}
    \def\@oddhead{}
    \def\@evenhead{}\let\@evenfoot\@oddfoot}
\def\draftdate{\number\month/\number\day/\number\year\ \ \ \hourmin }

\global\def\draftcontrol{0}
\def\citen#1{\if@filesw \immediate\write \@auxout {\string\citation{#1}}\fi%
\@tempcntb\m@ne \let\@h@ld\relax \def\@citea{}%
\@for \@citeb:=#1\do {\@ifundefined {b@\@citeb}%
    {\@h@ld\@citea\@tempcntb\m@ne{\bf ?}%
    \@warning {Citation `\@citeb ' on page \thepage \space undefined}}%
    {\@tempcnta\@tempcntb \advance\@tempcnta\@ne
    \setbox\z@\hbox\bgroup\ifcat0\csname b@\@citeb \endcsname \relax
    \egroup \@tempcntb\number\csname b@\@citeb \endcsname \relax
    \else \egroup \@tempcntb\m@ne \fi \ifnum\@tempcnta=\@tempcntb
    \ifx\@h@ld\relax \edef \@h@ld{\@citea\csname b@\@citeb\endcsname}%
    \else \edef\@h@ld{\hbox{--}\penalty\@highpenalty
    \csname b@\@citeb\endcsname}\fi
    \else \@h@ld\@citea\csname b@\@citeb \endcsname \let\@h@ld\relax \fi}%
\def\@citea{,\penalty\@highpenalty\hskip.13em plus.13em minus.13em}}\@h@ld}
\def\@citex[#1]#2{\@cite{\citen{#2}}{#1}}%
\def\@cite#1#2{\leavevmode\unskip\ifnum\lastpenalty=\z@\penalty\@highpenalty\fi%
  \ [{\multiply\@highpenalty 3 #1%
  \if@tempswa,\penalty\@highpenalty\ #2\fi}]}   %
\makeatother 
\catcode`\@=12

\def\be            {\begin{eqnarray}}
\def\bearl         {\begin{array}{l}}
\def\bearll        {\begin{array}{ll}}

\def\chii          {\raisebox{.15em}{$\chi$}}

\global\def\draftcontrol{0}

\def\ee            {\end{eqnarray}}
\def\eear          {\end{array}}

\newcommand\labl[1]{\label{#1}\ee \ifnum\draftcontrol=1
                   \mbox{ }\\[-12 mm]\query{#1}\\[5 mm] \fi}
\newcommand\Labl[2]{\label{#1#2}\ee \ifnum\draftcontrol=1
                   \mbox{ }\\[-12 mm]\query{#1#2}\\[5 mm] \fi}

\long\def\query#1{\hskip 0pt{\vadjust{\everypar={}\small\vtop to 0pt{\hbox{}%
     \vskip -13pt\rlap{\hbox to 49.0pc{\hfil{\vtop{\hsize=8pc\tolerance=6000%
     \hfuzz=.5pc\rightskip=0pt plus 3em\noindent#1}}}}\vss}}}}%

\newcommand\sect[1]{\section{#1}\setcounter{equation}{0}}

\def\g             {{\bf g}}

\def\d    {{\partial}}
\def\db   {{\bar{\partial}}}

\documentclass[12pt]{article}

\usepackage{amssymb,amsfonts,epsf,color,colordvi,fancybox}
\usepackage[dvips]{graphics} 
\begin{document}

\begin{flushright}  {~} \\[-1cm] {\sf LPTHE-03-26} \\[1mm]
{\sf October 2003} \end{flushright}

  \begin{center} \vskip 22mm
  {\Large\bf Discrete Torsion and WZW Orbifolds}
 \\[22mm]
 {\large Pedro Bordalo }\,
 \\[13mm]
 LPTHE, Universit\'e Paris VI~~~~{}\\
 4 place Jussieu\\
 F\,--\,75 252 Paris Cedex 05\\[5mm]
  \end{center} \vskip 23mm

\begin{quote}{\bf Abstract}
\\[3mm]
We propose a geometrical interpretation for the discrete torsion 
appearing in the algebraic formulation of quotients of WZW models
by discrete abelian subgroups.  
Part of the discrete torsion corresponds to the choice of 
action of the subgroup, yielding different quotient spaces.
Another part corresponds to the set of different choices of 
connection for the $H$ field in each of these spaces. 
The former is for instance used to describe generalized lens spaces $L_{(n,p)}$.
\end{quote}

\newpage

\sect{Introduction}

In perturbative string theory, space-time is built out of a 
conformal world-sheet theory.  It has therefore been
a constant theme to try to identify geometrically the features of the conformal field theory, 
as has been done most notably with boundary conditions on the worldsheet and D-branes.  
Another prominent example is that of orbifolding (see for example \cite{DHVW}, \cite{DVVV},
\cite{BHO}).   For instance, it has been shown in \cite{Vafa} that 
upon orbifolding {\it flat space} by some abelian, finite group $\Gamma$ one can include
phases in the twisted sectors of the theory, which correspond geometrically to introducing a 
non-trivial two-form field $B$, called discrete torsion.  In this note we extend this study to
a class of orbifolds of curved (bosonic) backgrounds.

We were motivated by the fact that the CFT describing string propagation in 
lens spaces $L_n=SU(2)/\mathbb{Z}_n$, where the left $SU(2)$ symmetry is preserved, 
is an element in a family of theories
corresponding to a particular choice of a parameter also called (algebraic) discrete torsion
\cite{KS}, \cite{BoWu}.  The question arose as to whether a different
choice of algebraic discrete torsion would also have a geometrical interpretation.
It is our goal here to describe the geometry of this family of CFTs.

To answer this question 
we will study more general orbifolds $M=G/\Gamma$ of compact, simple, simply connected groups 
$G$ by abelian, finite subgroups $\Gamma$ (whose action will be specified later). 
$\Gamma$ lies in some maximal torus of $G$ and is of the form
$\Gamma = \mathbb{Z}_{n_1} \times \ldots \times \mathbb{Z}_{n_r}$ where $r$ is the rank of $G$.
Depending on the action of $\Gamma$, the quotient space may be smooth or have some singularities.  
If $G/\Gamma$ is smooth, it inherits the left translational invariant metric $g$ and three-form 
$H$ from the covering $G$.  The quotienting can therefore be implemented on the WZW model 
describing $G$ at some level $k$.  Recall however that, for integrality of the Wess-Zumino term, 
the field $H$ is quantized, $H\in H^3 (G,k\mathbb{Z})$, so that it describes a (1-)gerbe over $G$ \cite{GaRe}. 
In order that the integrality of the Wess-Zumino term be preserved by the quotient, 
all $n_i$ must divide $k$.  This is not a restriction, since in comparing CFT with geometry one takes the 
semi-classical limit of $k$ large.  On the other hand, the theory may depend on the choice of connection
$B$ for the $H$-field \cite{Sharpe}.

We show that these same CFTs can be constructed using the simple current formalism \cite{KS}\footnote{In fact, 
the case of non-abelian $\Gamma$ can in principle be studied with tools 
introduced just recently in \cite{FRS} which are beyond the scope of this paper.}.  
This is a powerful tool which has been systematically studied for over a  
decade; we briefly review it in section \ref{cft}, and in particular we describe 
the algebraic discrete torsion. In section \ref{geometry} we review the construction of 
gerbes with connections on quotient spaces and show how they allow for inequivalent 
choices of $B$ fields. Using the K\"unneth formula 
we compare this freedom to the choice of CFT discrete torsion in the case of orbifolds
of WZW models.  We construct the partition functions for these quotient spaces 
and match them to the modular invariants derived in the section \ref{cft}; we find
that part of the discrete torsion corresponds to the choice of 
action of the subgroup, yielding different quotient spaces, and
another part corresponds to the different choices of $B$ field.

For some values of the discrete torsion, the correspondence above associates CFTs 
to target spaces $G/\Gamma$ having fixed points. We expect that these singularities are 
regulated by string theory in some way, but we leave the determination of the resolved geometry for 
later work. We conclude with some remarks on T-duality and D-branes on these spaces.

\sect{The CFT description of discrete torsion}\label{cft}

We start by building a rational conformal field theory corresponding to the sigma-model on 
the quotient $G/\Gamma$ with $\Gamma$ acting on the right.  We shall use 
the simple current formalism \cite{KS}, which we briefly review now.  
Given a rational CFT with chiral algebra ${\bf g}_k$, simple currents are those primary fields 
which have the property that they have an inverse under the fusion product, denoted $\star$.  A ubiquitous example is
the identity field itself. Furthermore, since the theory is rational, for any simple current $J$ there is
a positive integer $n_J$, called the order of $J$, such that $J^{n_J}=J\star \ldots \star J$ is the identity field.
The set of such primaries forms an abelian finite group for the fusion product, which is called
the center of the theory.  We are interested in those simple currents which have trivial monodromy
around all other fields, since these encode some ``bosonic'' symmetry of the theory.  Such 
simple currents form the effective center $\mathcal{Z}$, which is equivalently defined as
the subgroup of the center consisting of those simple currents $J$ for which 
$n_J\Delta_J \in \mathbb{Z}$ (where $\Delta_J$ is the conformal weight of $J$).

The partition function of a CFT is a bilinear in the characters of the irreducible highest 
weight representations
\begin{equation} \label{Z}
Z = \sum_{\lambda,\lambda'\ \mathrm{irreps} } Z_{\lambda \lambda'} \chi_\lambda \bar{\chi}_{\lambda'}
\end{equation}
It is said to be of {\it simple current type} if $Z_{\lambda\lambda'}$ is non-zero
only for $\lambda'= J\star \lambda$ for some primary $\lambda$ and some $J\in \mathcal{Z}$.  The consistent 
modular invariant partition functions of this type were classified in \cite{KS}, and they
constitute the vast majority of known rational CFTs. 

To build a modular invariant of simple current type one begins by choosing a simple current group 
$\mathcal{G} \subset \mathcal{Z}$.  Since $\mathcal{G}$ is a finite abelian group, it is of the form 
$\mathcal{G}=\mathbb{Z}_{n_1} \times \ldots \times\mathbb{Z}_{n_q} $.  
Then, picking a set of generators $J_i$ for $\mathcal{G}$ defines 
a $q\times q$ symmetric matrix of relative monodromies
$R_{ij} = Q_{J_i} (J_j) = \Delta(J_i) + \Delta(J_j) - \Delta(J_i \star J_j)$.  We define a $q\times q$ 
matrix $X$ whose entries are defined modulo integers and such that
its symmetric part is fixed by $X+X^t = R$. Its antisymmetric part is constrained
by $X_{ij} = x_{ij}/n_{ij}$, with $x_{ij}$ an integer and $n_{ij}=gcd(n_i,n_j)$.
The antisymmetric part of $X$ is called (algebraic) {\it discrete torsion}, and it is parametrized 
by $H^2(\mathcal{G},U(1))$ \cite{KS}.   Given a choice of discrete torsion, the modular invariant is given by 
\begin{equation}
Z = \sum_\lambda \sum_{J^{\vec{s}}\in\mathcal{G}} \left(\prod_{i=1} ^q 
    \delta_{\mathbb{Z}} (Q_{J_i} (\lambda)+X_{ij} s_j)\right)
    \chi_\lambda \bar{\chi}_{J^{\vec{s}}\lambda}  \label{modinv}
\end{equation}
where we have written the general element of $\mathcal{G}$ as $J^{\vec{s}}=\Pi_i J_i ^{s_i}$.
Here $\delta_{\mathbb{Z}}(a)=1$ if $a\in \mathbb{Z}$ and $0$ otherwise. 
So we see that the discrete torsion affects only the {\it off diagonal} ($s_i \neq 0$) terms 
in the modular invariant.  The left and right kernels of $X$ determine the extensions of the
left and right chiral algebras respectively.  Therefore, when they are different the modular
invariant is left-right asymmetric.

\bigskip

\paragraph{WZW models from Cosets:}
To describe the $G/\Gamma$ theory we start from the CFT  ${\bf g}_k /u(1)^r \otimes u(1)^r$, 
with $\g$ the Lie algebra of $G$, since the chiral $u(1)$ symmetries are not broken 
by the quotient (recall that $\Gamma$ is a subgroup of a maximal torus and that the level $k$ 
must be a multiple of the $n_i$).  To get acquainted with the procedure, we first recall how to 
recover the $G_k$ WZW (charge-conjugation) partition function as a modular invariant of the
${\bf g}_k /u(1)^r \otimes u(1)^r$ theory, following \cite{Gep87}.

Denote by $P$ the weight lattice of the finite dimensional Lie algebra $\g$ and by $Q$ its
root lattice.  From the decomposition of representations of $\g$ in terms of those of $u(1)^r$,
the characters of the irreducible highest weight representations of $\g_k$ decompose as
\begin{equation} \label{decomp}
\chi^{\g_k}_\Lambda (\tau,z) = \sum_{\mu \in P/kQ} \chi_{\Lambda ,\mu}(\tau) \chi_\mu ^{u(1)^r}(\tau,z)
\end{equation}
The generalized parafermion labels $(\Lambda,\mu)$ are subject to the selection rule $\Lambda-\mu \in Q$.
Furthermore, there are field identifications generated by outer automorphisms $\omega$ of $\g_k$
which descend to outer automorphisms of $u(1)^r_{2k}$.

The $G_k$ WZW (charge-conjugation) partition function is a simple current modular 
invariant with simple current group $\mathcal{G} \cong Q/kQ$.  For concreteness, in the following 
we shall mainly consider the case $\g_k = su(N+1)_k$.  In this case the rank $r$ is equal to $N$ and
$Q=\mathbb{Z}^N$.  
The parafermion primaries are vectors $(\vec{\jmath},\vec{m})$ which we write in a basis 
$\{ \vec{\lambda}_i\}$
for the weight lattice, e.g. $\vec{\jmath} = \sum_{i=1}^N j_i \vec{\lambda}_i$.  The 
parafermions are subject to the selection rules $m_i=j_i$ mod $2$ and to the field identifications
given by $(\vec{\jmath},\vec{m}) \sim (\omega{\vec{\jmath}},\vec{m}+k\vec{\lambda}_1)$ with
\begin{equation}
\omega(\vec{\jmath}) = \left( k- \sum_i j_i \right)\vec{\lambda}_1 + j_1 \vec{\lambda}_2
+ \ldots + j_{N-1}\vec{\lambda}_N
\end{equation}
In our notations both the parafermion labels $m_i$ and the $u(1)_{2k}$ labels, denoted
$m'_i$, are defined modulo $2k$.  The simple current group in this case is 
$\mathcal{G}\cong \mathbb{Z}_k ^N$, where the $i^{th}$ factor is generated by $J_i$ 
acting as $m_i \rightarrow m_i +2$ together with $m'_i \rightarrow m'_i +2$.

Now to define the modular invariant (\ref{modinv}) with simple current group 
$\mathcal{G}\cong \mathbb{Z}_k ^N$ we must specify the $N\times N$ matrix $X$.  For $N>1$
this allows for a choice of discrete torsion, i.e. the antisymmetric part of $X$ whose
entries are constrained to be of the form $x_{ij}/k$, for some integers $x_{ij}$.  To recover
the $G_k$ theory, i.e. to get the combinations of the characters of the form (\ref{decomp})
in the partition function (\ref{modinv}), we set the $x_{ij}$'s to zero.  Other choices
of $x_{ij}$'s will yield more complicated partition functions;  in this case it is not clear 
how to take the semi-classical limit of large $k$, since different scalings of the
$x_{ij}$ with $k$ yield different partition functions.  Therefore, it is not evident to 
assign a geometrical interpretation to non-zero values of these parameters.  In this note
we will set the $x_{ij}$'s to zero.

\bigskip

\paragraph{$G/\Gamma$ models from Cosets:}
The results in the previous paragraph suggest that the simple current group leading to the 
$G/\Gamma$ theory is $\mathcal{G} \cong \mathbb{Z}_k ^N \times \Gamma$.  There may be several
groups isomorphic to $\Gamma \cong \mathbb{Z}_{n_1} \times \ldots \times \mathbb{Z}_{n_N}$ in 
the effective center $\mathcal{Z}$, but without loss of generality we may assume that the
generators of $\Gamma$ act on directions of the maximal torus normal to each other
(with respect to the Killing form).  That means
that the $i^{th}$ generator of $\Gamma$, which we denote $W_i$, $i=1,\ldots,N$,
acts on the momenta as $m'_i \rightarrow m'_i + 2k/n_i$.
leaving all other quantum numbers invariant.  Here $n_i$ denotes the order of the current $W_i$, 
which may be one. Let us now consider the possible discrete torsion.  
From \cite{KS} we know that the (algebraic) discrete torsion for a theory with simple current 
group $\mathcal{G}$ is parametrized by
\begin{equation} \label{dtgroup}
\mathcal{D}(\mathcal{G}) \equiv H^2(\mathcal{G},U(1))
\end{equation}
We will see in section \ref{geometry} that for our case\footnote{Here we use interchangeably
the notations $\times$ and $\oplus$ for direct products of finite groups.}
\begin{equation} \label{dtgroup2}
H^2 (\mathbb{Z}_k ^N \times \Gamma,U(1)) = \mathbb{Z}_k ^{\frac{N(N-1)}{2}} \oplus
 \bigoplus_{i=1} ^N \mathbb{Z}_{n_i} ^N \oplus \bigoplus_{i<j} \mathbb{Z}_{gcd(n_i,n_j)}
\end{equation}
We can see this more directly by studying the antisymmetric part of the matrix $X$. 
For this, a word on notation is in order.  The matrix $X$ is written on a basis 
$\left(  J_i,W_j\right)$, where $i,j=1,\ldots,N$.  Since we want to keep the same index labelling $i$ 
for both sets of currents $J$ (the $\mathbb{Z}_k ^N$ generators) and 
$W$ (the $\Gamma$ generators)\footnote{This is because the indices $i$ are associated to the $U(1)^{(i)}$ 
subgroups of the chosen maximal torus.}, entries mixing $J_i$'s will be 
denoted $X_{ij}$, entries mixing $J$'s and $W$'s are denoted $Y_{ij}$ and entries mixing
$W$'s are denoted $Z_{ij}$.  Going back to the discrete torsion group (\ref{dtgroup2}), 
the first factor stands for the $x_{ij}$ mentioned in the previous paragraph.  Then,
to the entries $Y_{ij}$ we can add the numbers $y_{ij}/n_j$, because all
$n_j$ divide $k$.  Since the entries of the matrix $X$ are defined modulo integers, the 
$y_{ij}$ are parametrized by $\Gamma$.  This accounts for the $\mathbb{Z}_{n_i} ^N$
factors.  Finally, to the entries $Z_{ij}$ we may add $z_{ij}/gcd(n_i,n_j)$,
which are parametrized by $\bigoplus_{i<j} \mathbb{Z}_{gcd(n_i,n_j)}$.

So in our case the formula (\ref{modinv}) yields for general discrete torsion parameters
$(y_{ij},z_{ij})$ (recall we are setting $x_{ij}=0$)
\begin{equation}
Z(y_{ij},z_{ij})=\sum_{\vec{\jmath}}
 \sum_{\vec{s}} \left( \sum_{\vec{m},\vec{m}' (\vec{s})
      \atop{\vec{w},\vec{w}'(\vec{s})}}
 \chii_{\vec{\jmath}\vec{m}(\vec{s})}\chii_{\vec{m}'(\vec{s})}  
 \bar{\chii}_{j,\vec{w}(\vec{s})}   \bar{\chii}_{\vec{w}'(\vec{s})} \right)  \label{partition1}
\end{equation}
where we are writing the general simple current as 
$J^{\vec{s}}=\left( \Pi_i J_i^{s_i} \right)\left( \Pi_i W_i ^{\bar{s}_i} \right)$
together with $\vec{s}=(s_1,\ldots,s_N,\bar{s}_1,\ldots , \bar{s}_N)$ and
\begin{eqnarray}
m_i-m' _i & = & 
    \sum_{j}\left( y_{ij}+\delta_{i,j} \right) \frac{k}{n_j}\bar{s}_{j}\ \ \mathrm{mod}k 
         \label{cond1} \\
    m'_i &  = & (y_{ii}-1)s_i - \frac{k}{n_i} \bar{s}_{i} + 
       \sum_{j\neq i} \frac{z_{ij}}{n_{ij}}\frac{k}{n_i} \bar{s}_{j} 
       \ \ \mathrm{mod}\frac{k}{n_i}  \label{cond2}
\end{eqnarray}
and $w_i = m_i + 2s_i$, $w_i '= m_i ' + 2s_i + 2 \frac{k}{n_i} \bar{s}_{i}$.
Here we have used the notation $\chii_{\vec{m}'(s)} = \Pi_i \chii_{m'_i (s) } ^{u(1)^{(i)}}$
for the product of the $u(1)$ characters and $n_{ij}=gcd(n_i,n_j)$.

We will see in section \ref{general} that quotient $G/\Gamma$ with $\Gamma$ acting on the right
is described by the choice $x_{iI}=-1$, which preserves 
the left-moving ${\bf g}_k$ symmetry.  It is interesting to study
how properties of this family of theories depend on the discrete torsion.  This program has
been developed for general simple current modular invariants for what concerns boundary conditions,
see for instance \cite{FHSSW}.  In the next section we will focus on the geometrical interpretation 
of the discrete torsion.  

\sect{The geometrical description of discrete torsion} \label{geometry}

Let us consider for the moment a general background manifold $M$ with metric $g$ and three-form 
$H\in H^3 (M,\mathbb{Z})$.  To build a sigma-model with target space $M$, one specifies a gauge 
field (or connection) $B$ such that 
$dB=H$.  This connection $B$ is a collection of open sets $\mathcal{O}_i$ covering $M$, of 
2-forms $B^i$ on $\mathcal{O}_i$ which locally trivialize $H$ and such that $B^{(i)} -B^{(j)} = dA^{(ij)}$ 
are exact on 
$\mathcal{O}_{ij} = \mathcal{O}_i \cap \mathcal{O}_j $  -- this defines a Deligne cohomology
class of degree three\footnote{There are further conditions on the one-forms 
$A^{(ij)}$, for more details see \cite{GaRe} and references therein.}.  
Two connections are equivalent if and only if they differ 
by a connection of a line bundle, the choice of which is parametrized by the first cohomology group 
$H^1(M,U(1))$ \cite{GaRe}.  Furthermore, since in our case the three-form $H$-field
is quantized, the second cohomology group $H^2 (M, U(1))$ acts freely and transitively on 
the set of equivalence classes of connections.  
The role of $H^2 (M, U(1))$ for general backgrounds $M$ has been studied
by Sharpe (\cite{Sharpe} and references therein), in connection to the usual discrete torsion
introduced by Vafa \cite{Vafa}.
In general, $H^2 (M, U(1))$ is non-zero for $M$ a quotient of a simply connected manifold 
by some discrete group, which need not act freely.  

We now specialize to the case where $M$ is a quotient of a simple, simply-connected
compact Lie group $G$ by some action, indexed by $p$, of an abelian discrete subgroup $\Gamma$.
We denote the quotient by $M=G/\Gamma_p$.  Our results will apply to quotients of all compact Lie groups,
since they all have a simply connected Lie group as covering space.  As mentioned in
the introduction, $G/\Gamma_p$ inherits the translational invariant metric and $H$ field
from $G$, so that it is a target space for some CFT as long as the quotienting respects
the quantization condition on $H$.  For concreteness, and following Section \ref{cft}, most of the
calculations will be for $G=SU(N)$, but the generalization to $G$ a general simple, simply connected,
compact Lie group is straightforward.  

Before entering into more detail, it is useful to compare the group $H^2 (G/\Gamma_p, U(1))$
parametrizing the usual discrete torsion (\`a la Vafa) with the algebraic discrete torsion group
$\mathcal{D}(\mathbb{Z}_k ^N \times \Gamma)$ (\ref{dtgroup}).
First we notice that for $G$ simply connected, using the fact that $\pi_2 (G) = 0$, we have that 
$$
H^2(G/\Gamma_p,U(1))=H^2(\Gamma,U(1))
$$
even if the action of $\Gamma$ is not free (in which case the lhs stands for the $\Gamma$-equivariant
cohomology \cite{Sharpe}).  This allows us to work exclusively in terms of group cohomology.
Furthermore, from the short exact sequence 
$$0\rightarrow \mathbb{Z} \rightarrow \mathbb{R} \rightarrow U(1) \rightarrow 0$$
and using the fact that $H^*(\Gamma,\mathbb{R})=0$, we get that
\begin{equation} \label{iso}
H^l(\Gamma,U(1))=H^{l+1}(\Gamma,\mathbb{Z})
\end{equation}
for any positive integer $l$, and in particular we have 
$$\mathcal{D}(\mathcal{G}) = H^2(\mathbb{Z}_k\times\Gamma,U(1)) \cong 
H^3(\mathbb{Z}_k\times\Gamma,\mathbb{Z})$$  
To determine $H^*(\Gamma,\mathbb{Z})$ we use the K\"unneth formula (\cite{Brown})
which implies that if $X,Y$ are two finite abelian groups then
\begin{equation}  \label{kunneth}
H^*(X\times Y) = \left( H^*(X)\otimes_{\mathbb{Z}}  H^*(Y) \right) \times
                            Tor\left(H^*(X),H^{*+1}(Y) \right)
\end{equation}
We use here the standard mathematical notation that the sum of the degrees on the rhs is equal to
the degree on the lhs.   $Tor$ is a finite abelian group, which verifies 
$Tor(\mathbb{Z},\cdot) = Tor(\cdot, \mathbb{Z}) = 0 $ and also
\begin{eqnarray*}
Tor\left( \mathbb{Z}_n,\mathbb{Z}_m\right) = \mathbb{Z}_{gcd(n,m)}, \quad 
Tor\left( X , Y\times Z\right) = Tor\left( X , Y\right)\times Tor\left( X , Z\right)
\end{eqnarray*}
These properties will allow us to determine $H^*(\Gamma,\mathbb{Z})$.  Using (\ref{kunneth}) 
recursively for $\Gamma = \mathbb{Z}_{n_1} \times \ldots \times \mathbb{Z}_{n_N}$ we get for the
usual discrete torsion
\begin{equation} \label{u1}
H^2 (\Gamma,U(1)) \cong H^3(\Gamma,\mathbb{Z}) = \oplus_{i<j} \mathbb{Z}_{gcd(n_i,n_j)}
\end{equation}
Similarly, we compute the algebraic discrete torsion group to be (omitting the factor 
$\mathbb{Z}_k ^{\frac{N(N-1)}{2}}$)
\begin{equation}  \label{dtgroup3}
\mathcal{D}(\mathcal{G}) =  \bigoplus_i \mathbb{Z}_{n_i} ^N \oplus \bigoplus_{i<j} 
      \mathbb{Z}_{gcd(n_i,n_j)}
\end{equation}
This comparison suggests that the usual discrete torsion, that is the choice of connection $B$, 
is accounted for in the second factor of the CFT discrete torsion group 
$\mathcal{D}(\mathcal{G})$ (\ref{dtgroup})\footnote{That different choices of discrete
torsion can be related to inequivalent choices of a connection on a bundle
gerbe was also brought to my attention in discussions with Ch. Schweigert.}. 
Since this action exhausts the transformations compatible 
with the given background, we should expect that the CFT parameters encoded in the first factor 
$\Gamma^N$ will correspond to a change of background.  In order to study these parameters, 
we now consider the simplest example of the generalized lens spaces $L_{(n,p)}$.

\subsection{The example of lens spaces}\label{lens}

A generalized lens space\footnote{I thank G. Moore for bringing these spaces to my attention.} 
$L_{(n,p)}$ is a quotient of $SU(2)$ by the equivalence relation \cite{Brown}
\begin{equation}\label{equiv}
g \sim \omega^{\frac{p+1}{2}}\  g\  \omega^{\frac{p-1}{2}}
\end{equation}
where $\omega \in \mathbb{Z}_n \subset SU(2)$ is an element 
of order $n$.  Notice $p$ is defined modulo $n$, since $\omega^{n/2}$ is a central element
of order two.  Usual lens spaces correspond to the choices $p=\pm 1$.
In terms of Euler coordinates 
$g(\chi,\theta,\phi)=e^{i\frac{\chi}{2}\sigma_3}e^{i\frac{\theta}{2}\sigma_1}
                                                e^{i\frac{\phi}{2}\sigma_3}$
this action amounts to 
\begin{equation} \label{shift}
\chi\rightarrow \chi+ 2\pi \frac{p+1}{n}, \ \ 
\phi\rightarrow \phi- 2\pi \frac{p-1}{n}
\end{equation}  
If $n,p$ are not relatively prime, the action has a fixed circle $\chi+\phi=\mathrm{const}$.  

The sigma-model describing the usual lens spaces $L_{(n,1)}$ first appeared in \cite{GPS} as the bosonization 
of a two-dimensional supersymmetric CFT which appears in the near horizon limit of an extremal 
four dimensional black hole.  There the lens space was considered topologically as the Hopf fibration 
of $U(1)/\mathbb{Z}_n$ over the sphere $S^2$.  The $L_{(n,p)}$ spaces with $n,p$ relatively prime
are topologically very similar.  In particular, their fundamental group is 
$\pi_1 (L_n) \simeq \mathbb{Z}_n$ with the fiber being the non-trivial $S^1$; the maps in the 
different connected components of $\pi_0 (Map(S^1,L_n)) \simeq \mathbb{Z}_n$ 
differ by the winding number around the non-trivial $S^1$. 
Finally, their second cohomology group $H^2 (L_{(n,p)},U(1))$ is trivial, so that
there is only one class of connections $B$ for an integral $H$-field.  

To study string theory on $L_{(n,p)}$, we note that the action (\ref{shift}) breaks the 
$su(2)_k\times su(2)_k$ 
symmetry down to  a $u(1)_{2k} \times u(1)_{2k}$ symmetry, which means that the appropriate 
chiral algebra is $su(2)_k/u(1)_{2k}\times u(1)_{2k}$, with $k$ a multiple of $n$.  
Using the decomposition of (chiral) vertex operators
$$
V^{su(2)_k}=\sum_{m=0 \atop m=j(2)} ^{2k-1} V^{PF}_{jm} V^{u(1)_{2k}}_m
$$
and accounting for the action (\ref{shift}) on the left and right moving $u(1)$ fields, 
we can write a general vertex operator of the generalized lens
space CFT as\footnote{To simplify the notation we omit the superscripts.  
The vertex operators are recognizable by their labels.}
\begin{equation}\label{vertex2}
V[j,m,w,\bar{s}_1]= V_{jm}V_{m-(p-1)\frac{k}{n}\bar{s}_1}  
       \cdot \bar{V}_{jw}\bar{V}_{w-(p+1)\frac{k}{n}\bar{s}_1} 
\end{equation}
where $\bar{s}_1$ is a winding number.  Here the fact that $p$ is
defined modulo $n$ can be related to the the field identification in 
the parafermion sector $V_{jm} \sim V_{k-j,m+k}$.   Notice that the left
and right movers in (\ref{vertex2}) are related by simple currents of the
$su(2)_k/u(1)_{2k}\times u(1)_{2k}$ theory which form a group
$\mathcal{G}=\mathbb{Z}_k \times\mathbb{Z}_n$. 

To get the spectrum, we first impose level matching on the vertex
operators (\ref{vertex2}) (allowing for arbitrary lowering operators, which do not affect the level 
matching condition) and then impose modular invariance. 
In fact, due to the classification of simple current modular invariants \cite{KS}
we know there is only one partition function with sectors of the form (\ref{vertex2}). 
This is
\begin{equation}
Z\left( L_{(n,p)} \right) = \sum_{j=0} ^k
 \sum_{s_1,\bar{s}_1} \left( \sum_{m-m' = (p+1)\frac{k}{n}\bar{s}_1\atop{m+m' = 
                                              (p-1)(\frac{k}{n}\bar{s}_1+s_1) \bmod 2n}}
 \bar{\chii}_{jm}^{PF_k}\bar{\chii}_{m'}^{U(1)_k}  
 \chii_{j,m+2s_1}^{PF_k}\chii_{m'+2s_1+2\frac{k}{n}\bar{s}_1}^{U(1)_k}\right)  \label{partition2}
\end{equation}
which matches the partition functions (\ref{partition1}) for $p=y_{11}$; in particular, 
this parameter $p$ is parametrized by the discrete torsion group  
$\mathbb{Z}_n  = \mathcal{D} (\mathbb{Z}_k \times\mathbb{Z}_n) $.  
The matching of the partition functions written explicitely in terms of characters 
means that the CFTs are identical, so we have found that the different choices of the discrete torsion 
correspond geometrically to the {\it different target spaces} obtained by taking
the quotients (\ref{equiv}).
In particular, for $p=-1$ we recover the usual lens space $SU(2)/\mathbb{Z}_n$ \cite{GPS}
\begin{equation}
Z\left( L_{n} \right) = \sum_{j=0} ^k\chii ^{SU(2)_k} _{j}
 \left( \sum_{m-m' = 0 \bmod 2\frac{k}{n}\atop{m+m' = 0 \bmod 2n}}
 \bar{\chii}_{jm}^{PF_k}\bar{\chii}_{m'}^{U(1)_k}\right)  \label{partition}
\end{equation}

\subsection{The general case} \label{general}

We now consider the general case of a compact, simply connected Lie group $G$ and 
$\Gamma\cong\mathbb{Z}_{n_1} \times \ldots \times \mathbb{Z}_{n_N}$ an abelian 
subgroup of $G$, where some of the $\mathbb{Z}_{n_i}$ may be trivial.  
From our experience with the generalized lens spaces, we know that the discrete 
torsion parameters $y_{ij}$ of (\ref{cond1}) are related to the choice 
of equivalence relation, call it $p(y_{ij})$, which defines the target space
of the theory, denoted $G/\Gamma_p$.  Upon inspection of the selection rules for the 
left movers (\ref{cond1}) and for the right movers (just below that equation), the 
general form of the vertex operators is determined to be 
\begin{equation}
V_{\vec{\jmath},\vec{m}(\vec{s})}V_{\vec{m}'(\vec{s})}
\bar{V}_{\vec{\jmath},\vec{w}(\vec{s})}\bar{V}_{\vec{w}'(\vec{s})}
\end{equation}
with
\begin{eqnarray}
m_i-m_i' = \sum_{j} \left( y_{ij} + \delta_{i,j}  \right) \frac{k}{n_j} \bar{s}_{j}\ \ \mathrm{mod}k \\
w_i-w_i' = \sum_{j} \left( y_{ij} -\delta_{i,j} \right) \frac{k}{n_j} \bar{s}_{j}\ \ \mathrm{mod}k 
\end{eqnarray}
We find that the equivalence relation $p(y_{ij})$ which determines the target space 
of this theory is
\begin{equation} \label{equiv2}
g \sim  h_i^L\  g\  h_i ^R, \ \ \  i=1,\ldots,N
\end{equation}
where 
\begin{equation}
h_i^L = \omega_i ^{\frac{y_{ii}+1}{2}} \Pi_{j\neq i} \omega_j ^{y_{ij}/2}, \ \ \ 
h_i ^R = \omega_i ^{\frac{y_{ii}-1}{2}} \Pi_{j\neq i} \omega_j ^{-y_{ij}/2}
\end{equation}
and $\omega_i$ are elements of order $n_i$ along $N$ directions 
of the maximal torus so that they commute among themselves.  
Notice that for non-zero values of $y_{ij}$ this is an action of a subgroup 
of $\Gamma$, but not of the whole $\Gamma$. This relation generalizes (\ref{equiv})  
and accounts for the geometrical interpretation of the parameters $y_{ij}$.
Again, if $y_{ij}$ and $n_j$ are not relatively prime there will
be fixed point circles. 

Now given the set of allowed vertex operators, there will in general be a finite number of possible 
partition functions, corresponding to different choices of connections $B$ for the three-form $H$-field.
Recall that because the $H$-field is quantized, these are parametrized by the second cohomology
$H^2 (G/\Gamma_p,U(1))$.  Furthermore, since $H^3(G/\Gamma_p,\mathbb{Z})$ is without torsion, the
action of $H^2 (G/\Gamma_p,U(1))$ on a connection $B$ can be formally written in a way which 
resembles a globally defined gauge transformation \cite{GaRe}
\begin{equation}
B \rightarrow B + F  \label{gauge}
\end{equation}
where $F$ is a closed two-form on $G/\Gamma_p$.  
The transformation (\ref{gauge}) does not change the connection equivalence class if and only if 
$F \in  H^2 (G/\Gamma_p, \mathbb{Z})$ so we recover the action of 
$H^2(G/\Gamma_p,U(1))$ on the set of equivalence classes of connections.  
These considerations are only formal because $H^2(G/\Gamma_p,U(1))$ and 
$H^2(G/\Gamma_p,\mathbb{Z})$ in general have torsion, 
and only their non-torsion elements can be related to the de Rham cohomology.  In any case, 
if the target manifold has non-trivial second cohomology group, we may expect that the 
sigma-model path integral depend on the second homology of the embedding of the worldsheet 
$\iota(\Sigma)$ in $G/\Gamma_p$.  Indeed, as described in \cite{Sharpe} the group 
$H^2 (G/\Gamma_p,U(1))$ acts by multiplying the path integrand by a phase
\begin{equation}  \label{b}
e^{iS} \rightarrow e^{i\int_C F_{ij} \d X^i \db X^j}e^{iS}
\end{equation}
where $C$ is the 2-cycle in $G/\Gamma_p$ wrapped by $\iota(\Sigma)$.  
Let us see how this may affect the spectrum.

Denote by $U(1)^{(i)}$ the subgroup of the maximal torus containing the factor 
$\mathbb{Z}_{n_i}$ of $\Gamma$. Multiplying by the phase (\ref{b})
will shift the left and right $u(1)$ currents along both the $i$ and the $j$ directions, by
$j^i \rightarrow j^i+F_{ai}\d X^a$ and $\bar{\jmath}^i \rightarrow \bar{\jmath}^i-F_{ai}\db X^a$
respectively (summation over $a$ intended).  This means the corresponding left and right moving fields 
will shift as
\begin{eqnarray} \label{shifts}
m'_i \rightarrow m'_i+F_{ai}\frac{k}{n_i} \bar{s}_a \nonumber \\
w'_i \rightarrow w'_i-F_{ai}\frac{k}{n_i} \bar{s}_a 
\end{eqnarray}
where $\bar{s}_a$ is the winding number.  Recall that the phase (\ref{b}) is a $n_{ij}^{\mathrm{th}}$ 
root of unity, where $n_{ij}=gcd(n_i,n_j)$, so $F_{ij}=z_{ij}/n_{ij}$ with $z_{ij}$ an integer.
Thus the action of  $H^2 (\Gamma,U(1))\cong \bigoplus_{i<j} \mathbb{Z}_{gcd(n_i,n_j)}$ on 
the spectrum indeed corresponds to the discrete torsion shift $z_{ij}$ in (\ref{cond2})
(this is essentially as described by Vafa \cite{Vafa}).  The simple current ``coordinates''
$\bar{s}_i$ are then interpreted as winding numbers.

So we have shown that where both sides are defined, discrete torsion on the CFT side and on 
the geometric side are equivalent.  It would be interesting to find out whether
this holds true even when one side is not yet well known.  For instance, it is not always clear what 
a different choice of simple current group $\mathcal{G}$ corresponds to 
geometrically, but it should allow for the discrete torsion which is present in the CFT side.  
On the other hand, we may consider backgrounds whose topology is known but where the corresponding 
CFT is not well known, such as more general quotients by abelian groups, obtained by generalizing 
(\ref{equiv2}) to an equivalence relation with two parameters, a $p_\mathrm{left}$ and a $p_\mathrm{right}$.
Other more general situations would be for instance quotients by non-abelian 
groups or quotients of more complicated groups such as products of groups or 
non-compact groups.  We expect to find in the corresponding CFT a free parameter 
which corresponds to the geometric freedom of choice of connection.  

\subsection{Remarks}

The T-dual version of these models can be easily studied in the CFT approach.  
For the choice of discrete torsion which preserves the chiral symmetry, i.e. for $p_i = \pm 1$, 
T-dualizing along a $U(1)$ direction where a $\mathbb{Z}_{n_i}$
acts results in a theory where $\mathbb{Z}_{n_i}$ is replaced by $\mathbb{Z}_{k/n_i}$.  This 
was used extensively in \cite{MMS},\cite{BoWu} to study B-type branes in lens spaces.

However, for any different choice of discrete torsion, T-duality will yield a quotient 
of $G$ which, though preserving the left and right $u(1)$ symmetries, is not of the form
(\ref{equiv2}).
Indeed, in general, $\Gamma$ changes not only by replacing $\mathbb{Z}_{n_i}$ 
by $\mathbb{Z}_{k/n_i}$ but also by changing the generator of the diagonal $\mathbb{Z}_k$.
Finding the geometrical interpretation of such simple current modular invariants
would give more information on the exact T-duality transformations of a curved background.

\medskip

We should note that the group $H^1 (M,U(1))$  which parametrizes the gauge choices for 
the connection $B$ (cf. section \ref{geometry}) does affect the open string theory, since
its action amounts to introducing Wilson lines in the background.  To see this, let us 
first consider the symmetric boundary states in simple current CFTs \cite{FHSSW}.
All choices of discrete torsion appear on the same footing, so we can describe easily the 
boundary blocks, the boundary states and their annuli amplitudes in all the models 
considered above. 
In fact, only fractional boundary blocks, which appear whenever some $n_i$ is even,
depend on the discrete torsion. Let $D$ be the worldvolume of some fractional brane.
Then $H^1 (M,U(1))$ acts (projectively, via $H^1 (D,U(1))$) by transforming
the fractional branes with worldvolume $D$ into each other.  


\medskip

{\bf Acknowledgements.}
It is a pleasure to thank Gregory Moore for helpful discussions on generalized lens spaces 
and group cohomology.  I would also like to thank Christoph Schweigert, Pierre Vogel and 
Volker Braun for discussions on group cohomology, and Giuseppe d'Appollonio and 
Sylvain Ribault for many helpful comments.  

The author is supported by the grant SFRH/BD/799/2000 of FCT (Portugal).


\end{document}